\documentclass[12pt,titlepage]{article}
\usepackage{graphicx}
\usepackage{amssymb}
\usepackage{color}

\font\grande=cmr10 scaled \magstep4
\font\medio=cmr10 scaled \magstep2
\outer\def\beginsection#1\par{\medbreak\bigskip
      \message{#1}\leftline{\bf#1}\nobreak\medskip
\vskip-\parskip
      \noindent}

\def\laq{\raise 0.4ex\hbox{$<$}\kern -0.8em\lower 0.62
ex\hbox{$\sim$}}
\def\gaq{\raise 0.4ex\hbox{$>$}\kern -0.7em\lower 0.62
ex\hbox{$\sim$}}
\newcommand{\be}[1]{\begin{equation}}
\newcommand{\ee}{\end{equation}}
\newcommand{\eeqn}{\end{eqnarray}}
\newcommand{\bd}{\begin{displaymath}}
\newcommand{\ed}{\end{displaymath}}
\def\beq{\begin{equation}}
\def\eeq{\end{equation}}
\def\bea{\begin{eqnarray}}
\def\eea{\end{eqnarray}}
\def\bean{\begin{eqnarray*}}
\def\eean{\end{eqnarray*}}

\def \ra {\rightarrow}

\def \ra {\rightarrow}

\begin{document}
\bibliographystyle {unsrt}
\titlepage

\vspace{15mm}
\begin{center}
{\grande
ADM,  BMS, and some puzzling interconnections }
\vspace{15mm}

\vspace{6mm}
 \large{ Gabriele Veneziano}
\vspace{6mm}

{\sl Theory Department, CERN, CH-1211 Geneva 23, Switzerland} 

{\sl and} 

{\sl Coll\`ege de France, 11 place M. Berthelot, 75005 Paris, France} 

\end{center}

\vskip 2cm
\centerline{\medio  Abstract}
\vskip 5mm

\noindent
The precise connection between the ADM and BMS formalisms is still far from being fully understood. It leads superficially to some puzzles whose resolution can provide new interesting physical insights. One example concerns the claimed local-in-angle conservation of energy in a gravitational scattering process, whose physical meaning I will try to clarify in an explicit example. As my main topic, I will sketch my own understanding of how a recent  saga on the definition of angular momentum, its loss, and its radiation, appears to have found its way to a happy ending by appealing to some non-trivial connections between ADM and BMS quantities in asymptotically flat and stationary  spacetimes.
\vspace{5mm}

\vfill 
\begin{flushleft}
\end{flushleft}
\newpage

\section{Meeting Stanley, ADM, BMS}

I do not remember when exactly I bumped into the ADM formalism in General Relativity (GR). Certainly not during my undergraduate or graduate education. At the time (mid sixties) Particle Physics and General Relativity were  two non-communicating areas of theoretical physics. 

Probably I met Stanley before meeting ADM. 
It was soon after my arrival as a post-doc at MIT in the fall of 1968, i.e. only a short while after my paper on the dual amplitude had appeared. I got invited to give talks at several Institutes and one of the very (if not the) first was Brandeis University  in the greater-Boston area. 
Stanley was my host and I still remember his kind and flattering words. Possibly during some chats over coffee, I asked him what he was working on. In answering he may have  mentioned his  Hamiltonian approach to GR with Arnowitt and Misner  \cite{Arnowitt:1959ah, Arnowitt:1960es} (see \cite{Arnowitt:1962hi} for an easily accessible account of the original work) prompting me to get acquainted with it. More likely, I only did so later on, when I decided to study GR and cosmology in Weinberg's book, motivated by its chapter on early cosmology within theories, such as the dual resonance model, exhibiting a  Hagedorn spectrum and temperature.

I remember, instead, how I got acquainted with Bondi-Sachs (BS) coordinates, Bondi masses \cite{Bondi:1960jsa, Sachs:1962wk} (see \cite{Madler:2016xju} for a review) and their relation to ADM mass. It was thanks to Grisha Vilkovisky when, during the Academic year 1992-1993, he visited CERN and we worked together on the gravitational $S$-matrix. Together with two younger collaborators we managed to express the massless gravitational S-matrix in terms of past and future Bondi masses \cite{Fabbrichesi:1993kz}, a topic I want to come back to sooner or later. 

In this brief and modest contribution to the memory of a great physicist and dear friend I will describe some examples of the interconnection between the ADM and BS formalism as well as with the BMS asymptotic symmetry \cite{Bondi:1962px}, \cite{Sachs:1962zza} (see again \cite{Madler:2016xju} for a review) underlying the latter. ADM quantities  refer to the behaviour of the metric at spatial infinity $i^0$, while BS (or BMS) quantities refer to past or future null infinity, ${\cal I}^{\mp}$. Continuity arguments suggest that, for each physical observable, the $ v \to + \infty$ limit (of BS's advanced time $v$) and the $ u \to - \infty$ limit (of BS's retarded time $u$) should coincide and should be both given by the corresponding ADM quantity. 

I will present two examples of how this kind of reasoning bears fruit. In the first example, it clarifies the meaning of a claimed \cite{Strominger:2013jfa} local  (i.e. angle-by-angle) energy conservation by gravitational scattering. In the second, I will give my own understanding on how connecting ADM and BMS quantities may solve the long-standing issue of how to relate (and distinguish) angular momentum loss and radiation in GR.

In order to make the presentation sufficiently self contained, I will start with a schematic recap of BS and BMS formalism. Here, and when I will report on other people's work, I claim no credit, of course, and also take full responsibility for any possible misquoting.

 \section{Bondi-Sachs coordinates}

 In order to describe gravitational  (as well as other massless) radiation it is convenient to use an adapted coordinate system. 
 Having in mind a scattering process in which some (massive and/or massless) particles come in, interact (and possibly emit radiation) during a finite lapse of time,   such adapted coordinates are known as incoming (or advanced) and outgoing (or retarded) Bondi-Sachs (BS) coordinates \cite{Bondi:1960jsa}, \cite{Sachs:1962wk} depending on whether one wants to describe the far past or the far future of the process.  Concentrating for the moment on the latter, BS coordinates ($u, r, \theta^A$) are characterised by a line element of the form:
 \beq
 ds^2 = - U e^{2\beta} du^2  - 2  e^{2\beta} du dr +  \gamma_{AB}( r d \theta^A + U^A du) ( r d \theta^B + U^B du) ~;~  \det  \gamma_{AB} = \sin^2 \theta\, ,
 \label{BS}
 \eeq
 where, by their definition in (\ref{BS}),  $u$ is  an exactly null retarded-time coordinate and $r$ is the so-called area distance. Finally, $ \theta^A = ( \theta, \phi)$ are two polar coordinates on the unit 2-sphere.
 
 By definition, asymptotically flat spacetimes  admit an expansion of (\ref{BS}) in  inverse powers of $r$:
 \bea
 & U =  1 - \frac{2 G {\cal M}}{r} +{\cal O}(\frac{1}{r^2}) ~;~ \beta = {\cal O}(\frac{1}{r^2}) ~;~  U^A = \frac{D_B C^{AB} }{2 r } + \frac{2 G {\cal N}^A + {\cal O}(C^2) }{3 r^2 }+{\cal O}(\frac{1}{r^3}) \nonumber \\
 &\gamma_{AB}= \Omega_{AB} + \frac{C_{AB}}{r} + {\cal O}(\frac{1}{r^2}) ~~ ({\rm with} ~\Omega^{AB} C_{AB} =0) \, ,
 \label{BSexp}
 \eea
 where $\Omega_{AB}$ is the standard round metric on the unit 2-sphere and $D_A$ is the covariant derivative compatible with it. The asymptotic metric (\ref{BSexp}) thus depends on just five functions ($ {\cal M}, C_{AB},  {\cal N}^A$) of the three coordinates $(u, \theta^A)$~\footnote{Equations similar to (\ref{BS}), (\ref{BSexp}) can be defined near past null-infinity. The incoming (or advanced) coordinates are usually denoted by $(v, r, \theta^A)$ although all four of them should be distinguished from $(u, r, \theta^A)$.}. 
 
  ${\cal M}$ is the so-called mass aspect, while the two-vector ${\cal N}^A$ is called the angular momentum aspect\footnote{Its normalization and the precise terms ${\cal O}(C^2)$ added to its definition in (\ref{BSexp}) differ in the literature. Here we follow the conventions of \cite{Riva:2023xxm} (which are essentially those of \cite{Flanagan:2015pxa} to which we refer for further details).}. Finally, the symmetric traceless tensor $ C_{AB}$ is the shear tensor connected to the two polarizations of gravitational waves in four spacetime dimensions. Out of $ C_{AB}$ one can construct the so-called news tensor $N_{AB} \equiv \partial_u C_{AB}$ which, as we shall discuss below, carries information about the energy, linear and angular momentum carried by the gravitational waves\footnote{From their definition in (\ref{BSexp}) ${G \cal M}$ and $C_{AB}$ have dimensions of a length (having set $c=1$ throughout) while ${G \cal N}^A$ has dimensions of a squared length if ${\cal M}$ and ${\cal N}^A$ have dimensions of mass and angular momentum. In the following we shall often work with geometric units where one also sets $G=1$.}. 
  
   Einstein's equations (here written for simplicity in vacuum, but source terms can be easily added) tell us how $ {\cal M}$, and ${\cal N}_A$ evolve in $u$ once the shear tensor, and thus also the news tensor, are given. One finds:
\bea
 &\partial_u {\cal M}    =  - \frac{1}{8} N_{AB} N^{AB}  +  \frac14 D_AD_B  N^{AB}  \, , \nonumber \\
 & \partial_u {\cal N}_A   = D_A {\cal M} + \frac14 D_BD_AD_C C^{BC} - \frac14 D^2D^B C_{AB} + \frac14 D_B(N^{BC} C_{CA}) + \frac12 C_{AB}D_CN^{BC}\, .
 \label{eveq}
	\eea

 \section{BMS transformations}

 The metric form (\ref{BS}) does not fix completely the coordinate system. The residual (gauge) freedom essentially defines the group of BMS transformations \cite{Bondi:1962px}, \cite{Sachs:1962zza} (see again \cite{Madler:2016xju} for a review). It consists of  a global Lorentz group (rotations and boosts) and of ``super-translations", a generalization of ordinary translations that includes a $\theta^A$-dependent shift of $u$, accompanied by suitable transformations  of $r$ and $\theta^A$. More specifically, the leading-order infinitesimal transformations are:
 \beq
 u \ra u + \alpha (\theta^A) + \frac12 u D_A Y^A ~;~ r \ra r \left(1- \frac12 D_A Y^A \right) +  {\cal O}(r^0)  ~;~ \theta^A \ra \theta^A  + Y^A  +  {\cal O}(r^{-1})\, ,
 \label{BMS}
 \eeq
 where $Y^A(\theta^B)$ represents the Lorentz group by taking the form:
\beq
 Y^A = D^A \chi + \epsilon^{AB} D_B \kappa ~, \, {\rm with} ~ (D^2 +2)\chi  =  (D^2 +2)\kappa = 0\, .
 \label{Lg}
 \eeq
 The latter equations imply that $\chi$ and $\kappa$ are $l=1$ spherical harmonics representing the three generators of boosts and the three generators of rotations,  respectively.
 
 On the other hand, $\alpha$ is an arbitrary (regular) function of the angles. If we write it as
 \beq
 \alpha  =  \alpha_0 + \sum_{i=1}^3 \alpha_i n^i(\phi^A) + \tilde{\alpha}(\phi^A)~;~ n^i = (\sin \theta \cos \phi,  \sin \theta  \sin \phi, \cos \theta)  \, ,
  \label{St}
 \eeq
   $ \alpha_i$ and $ \alpha_0$ generate ordinary space and time translations (thus completing with $Y^A$ the full Poincar\'e group), while the higher harmonics ($l \ge 2$) in $\tilde{\alpha}(\phi^A)$ generate the so-called proper super-translations, the real novelty of the BMS group.

 Of much importance for what follows is the way the various entries of the BS metric transform under the BMS group and, in particular, under an infinitesimal super-translation defined by setting $Y^A=0$ in (\ref{BMS}). One finds:
 \bea
& \delta C^{AB} =  \alpha ~ N^{AB}  - 2 D^A D^B \alpha  + \Omega^{AB} D^2 \alpha~;~  \nonumber \\
& \delta {\cal M} = \alpha ~\partial_u {\cal M} + \frac14  N^{AB} D_A D_B \alpha + \frac12  D_A \alpha D_B N^{AB} ~;~  \nonumber \\
& \delta {\cal N}_A =  \alpha ~\partial_u {\cal N}_A  + 3 {\cal M}  D_A \alpha -
 \frac34  D_B \alpha (D^B D^C C_{CA} - D_A D_C C^{BC}) + \frac34 C_{AB} N^{BC} D_C \alpha\, .
 \label{sttransf}
 \eea

 In eqns. (\ref{sttransf}) the first terms on the r.h.s. correspond to the naive transformation of a scalar quantity. Additional terms are present for the variations of $C^{AB}$, $ {\cal M}$, and ${\cal N}_A$. For $\delta C^{AB}$ the extra terms are $u$-independent and only sensitive to the $\tilde{\alpha}$ piece in (\ref{St}). They  are also absent in $\delta {\cal M} $ in non-radiative regions where  $N_{AB}=0$. Instead, even in those regions,  $ \delta {\cal N}_A $ has additional contributions controlled by $C_{AB}$ and sensitive to the full  $\alpha$ of (\ref{St}), including the ordinary translation part.

 \section{BMS charges and fluxes} 

 In analogy with what happens with symmetries in flat spacetime one can associate charges with the generators of  BMS transformations although, as discussed in \cite{Wald:1999wa}, they are not conserved in the usual sense. Those associated with super-translations are called supermomenta and are defined  (as a function of $u$ and a functional of $\alpha$) by:
 \beq
P( u; \alpha) = \int_u  \alpha(\theta^A) {\cal M}(u, \theta^A) d^2 \Omega ~;~  d^2 \Omega \equiv  \frac{1}{4 \pi} d \theta \sin \theta d \phi\, , 
\label{SM}
\eeq
where the integral is over a two-sphere at the constant specified value of $u$. 
 
 Similarly, those associated with the Lorentz generators encoded in $Y^A$ are the Lorentz charges for which there are different definitions in the literature satisfying different requirements. They all contain the same combination of the angular momentum and mass aspects, but differ by additional shear-dependent pieces. Following  \cite{Flanagan:2015pxa} and  \cite{Riva:2023xxm} we adopt the original Dray-Streubel definition  \cite{Dray:1984rfa}:
  \beq
J(u;Y^A) = \frac12 \int_u Y^A(\theta^B) \left({\cal N}_A (u, \theta^B) - u D_A  {\cal M} - \frac{1}{16} D_A (C_{BC}C^{BC}) - \frac14 C_{AB}D_CC^{BC} \right) d^2 \Omega \, .
\label{LCh}
\eeq

 \subsection{Energy, momentum, and supermomenta evolution}
 
  Let us now discuss, separately, the (non) conservation laws for each type of charges (see \cite{Ashtekar:2024stm}  for a recent review of a coordinate-independent  formulation of these laws).
 
 For any given $\alpha$ the $u$-dependence of $P( u; \alpha)$ can be obtained by differentiating $ {\cal M}$ inside the integral (\ref{SM}) and by inserting the first of (\ref{eveq}) with the result:
  \beq
\partial_u P( u; \alpha) = - \frac{1}{8} \int_u  \alpha(\theta^A)  \left( N_{AB} N^{AB}  -2 D_AD_B  N^{AB}\right)   d^2 \Omega \, .
\label{SMev}
\eeq

A particular case of (\ref{SMev}) is obtained by taking $\alpha$ to represent ordinary space-time translations (i.e. by setting $\tilde{\alpha}=0$ in (\ref{St})). 
We thus define the Bondi mass $M$ and 3-momentum $P^i$ by:
\beq
M(u) =  \int_u {\cal M}(u, \theta^A) d^2 S~;~  P^i(u) =  \int_u {\cal M}(u, \theta^A) n^i d^2 S\, .
\label{B4M}
\eeq
For $M$, in particular, we  get the celebrated Bondi energy-loss formula\cite{Bondi:1960jsa}:
\beq
\label{Eloss}
  \frac{d M}{du}   =  - \frac{1}{8} \int_u N_{AB} N^{AB} d^2 S \, ,
\eeq
 since the second term in (\ref{SMev}) integrates to zero. The (semi)negative-definite term on the r.h.s. of (\ref{Eloss}) is interpreted as the energy lost to gravitational radiation.
 Integrating by parts, and using  the  identity  (see e.g. \cite{Bonga:2018gzr}) $D_AD_B n^i  = - n^i \Omega_{AB} $,  we also obtain a similar equation for the 3-momentum flow;
 \beq
\label{Ploss}
  \frac{d P^i(u)}{du}   =  - \frac{1}{8} \int_u N_{AB} N^{AB} n^i d^2 S \, ,
\eeq
 while for a generic super-momentum the last term in (\ref{SMev}) cannot be neglected. 
 
Note that in (\ref{SMev})  only the news tensor appears so that the (super)momenta loss formulae are invariant under BMS transformations. As we shall see this will be a crucial difference when compared to the case of Lorentz charges.

Before moving on to those let us make a digression. The Bondi mass\footnote{Here and in the Appendix we add an extra $\pm$ label to distinguish  $M(u, \theta^A)$ from $M(v, \theta^A)$. Elsewhere, unless it may generate confusion, we omit it.} $M_+(u)$ represents the energy that is still in the (finite distance) system at retarded time $u$.
As we take the limit $u \to - \infty$ we expect $M_+(u)$ to converge to the ADM mass, which is defined in terms of the metric at spatial infinity.

As already mentioned, one can consider incoming Bondi-Sachs coordinates $(v, r, \theta^A)$ describing data near past null infinity and  define, in the large-$r$ limit, a Bondi mass aspect ${\cal M}_-(v, \theta^A)$ and  a Bondi mass $M_-(v)$ representing the energy that has entered the system by the advanced time $v$ and claim that $M_-(v = + \infty)=M_{ADM}$. In other words, the  infinite past of future null infinity and the infinite future of past null infinity are both identified with spatial infinity, a single point in the Carter-Penrose diagram. The  limits of the two Bondi masses must  therefore coincide and they are indeed given by the corresponding  ADM quantity.

It has been argued \cite{Strominger:2013jfa} that not only $M_{\pm}$, but also ${\cal M}_{\pm}$, satisfy the above-mentioned matching at spatial infinity provided the two sets of angles in ${\cal M}_{+}$ and ${\cal M}_{-}$  are antipodally identified. This is sometimes referred to as a statement of energy conservation at each angle, a statement that looks at odds with the well established  phenomenon of gravitational deflection. In Appendix A we show how this apparent paradox gets explicitly solved  \cite{GVunp} at lowest order in the deflection angle, and refer to the literature for arguments reaching the same conclusion in more general situations.
 
 \subsection{Lorentz-charge evolution and the puzzle of angular momentum loss}
 
 Let us now try to proceed with the $u$-evolution of Lorentz charges following the way we used for supermomenta. We differentiate (\ref{LCh}) to get:
  \beq
\partial_u J(u;Y^A) = \frac18 \int_u Y^A(\theta^B) \left(\partial_u {\cal N}_A -  D_A  {\cal M} - u D_A  \partial_u {\cal M} + \dots \right) d^2 \Omega \, ,
\label{LCh}
\eeq
 where the dots represent terms bilinear in the shear and the news. Inserting now the second of eqns. (\ref{eveq}) and after quite some algebra one arrives at 
  \bea
&\partial_u J(u;Y^A) =\frac18  \int_u d^2 \Omega Y^A \left(N^{BC} D_A C_{BC}-2 D_B(N^{BC} C_{AC}) + \frac12 D_A (N^{BC} C_{BC})  - \frac12 u D_A X\right) \, ;\nonumber \\
& X \equiv  (N^{BC} N_{BC} - 2 D_BD_C N^{BC})  \, .
\label{LCh1}
\eea
 The important point here is that, unlike $X$, the rest of what appears in the Lorentz charges flow is {\it not} invariant under (proper) super-translations because of the presence of $C^{AB}$. This is the (in)famous Bondi-frame dependence of the (Bondi) angular momentum flow: it is so strong that in some processes it can even affect the order in $G$ at which the flow of $J$ starts. 
 
 Consider for instance the physically interesting case of the gravitational collision of two Schwarzschild black holes \cite{Veneziano:2022zwh}. The mass aspect, angular momentum aspect, and shear, all occur already at ${\cal O}(G)$ (one-graviton exchange), i.e. before any radiative process kicks in. The news, instead, start at ${\cal O}(G^2)$. Reintroducing the appropriate powers of $G$ in  (\ref{Eloss}), (\ref{LCh1}), one finds that the flow of energy starts at ${\cal O}(G^3)$, as expected\footnote{This parallels the well-known fact that, in QED, photon emission in an $e^+ e^-$ collision starts at ${\cal O}(\alpha^3)$.}, while the flow of angular momentum starts, in general,  one order of $G$ lower, because of the replacement of one $N_{AB}$ by a $C_{AB}$ in the respective flux formulae. There is only one Bondi gauge in which both  flows start at ${\cal O}(G^3)$ as expected for real-graviton emission: it is  the so-called canonical Bondi gauge (also known as the ``good-cut prescription") in which the initial shear vanishes so that
\beq
C_{AB}(u) = \int_{-\infty}^u du' N_{AB}(u') = {\cal O}(G^2)\, .
\label{CGshear}
\eeq
The  (naive) conclusion  thus appears to be that the loss of angular momentum has to be computed in the canonical Bondi gauge.  This, however, raises a different puzzle\cite{Veneziano:2022zwh}.

According to Bini and Damour  \cite{Bini:2012ji} there is a linear-response formula giving the back-reaction on the two body system (e.g. on the deflection angle) from  energy and angular momentum loss. Such a back reaction can also be computed by more direct methods \cite{DiVecchia:2021ndb, DiVecchia:2021bdo, Herrmann:2021tct, Bjerrum-Bohr:2021din, Brandhuber:2021eyq}  and does start at ${\cal O}(G^3)$\footnote{Including the back-reaction at  ${\cal O}(G^3)$ is also of crucial importance for restoring agreement \cite{DiVecchia:2020ymx} (see also \cite{Bern:2020gjj}) between the 3PM conservative calculation of \cite{Bern:2019nnu} and the old ultrarelativistic result of \cite{Amati:1990xe}.} which is consistent with the linear response  of \cite{Bini:2012ji} provided the energy or the angular momentum loss (or both)  start already at ${\cal O}(G^2)$. However, since the energy loss is unambiguously ${\cal O}(G^3)$, the only possibility is for the $J$-loss to be ${\cal O}(G^2)$. This  is indeed the case in the computation in \cite{Damour:2020tta} which, therefore,  cannot be consistent with computing the loss of $J$ from (\ref{LCh1}) in the canonical frame. 

This was the puzzle pointed out in  \cite{Veneziano:2022zwh}. In the next subsection I will summarise its (partial?) resolution  given there, and then present (my own understanding of) some subsequent developments possibly opening the way to solving completely this subtle issue via a BMS-ADM connection.

\subsection{Solving the $J$-loss  puzzle through a BMS-ADM connection?}

The physical reason given in \cite{Veneziano:2022zwh} for choosing the canonical gauge when calculating the radiated angular momentum  is that in this gauge, and only in this gauge, the Bondi angular momentum at $u=-\infty$ coincides with the ADM angular momentum. This statement was proven in \cite{Ashtekar:1979iaf} for stationary spacetimes. Soon after, the reasoning was generalized \cite{Ashtekar:1979xeo} to show that the past limit of the Bondi 4-momentum is the ADM 4-momentum for general radiating spacetimes. The two arguments can be combined to show that the past limit of the Bondi angular momentum computed in the canonical gauge equals the ADM angular momentum  also in radiating spacetimes \cite{Ashtekar:2019rpv, Ashtekar:2023wfn, Ashtekar:2023zul} provided the process is sufficiently stationary in the asymptotic past. The same conclusion was reached  by working in a Hamiltonian formalism through the addition of ``logarithmic supertranslations" (see  \cite{Fuentealba:2023syb} and references therein). This allows to define supertranslation-invariant Lorentz generators as part of an enlarged BMS algebra.
 
 It looks physically mandatory to count  the loss of angular momentum out of its initial ADM value.  There is also no doubt that gravitational radiation (seen as graviton emission) starts at ${\cal O}(G^3)$ in a purely gravitational collision so that the loss of angular momentum through radiation should be computed via (\ref{LCh1}) in the canonical Bondi gauge. The only way to reconcile these two statements with a {\it total} loss of angular momentum  ${\cal O}(G^2)$ is that the latter is {\it not} entirely due to radiation and thus cannot  be computed from (\ref{LCh1}) by sticking, at all $u$, to the canonical gauge.

It was argued in \cite{Veneziano:2022zwh} that there is another gauge relevant for the scattering problem.
 It is fixed by the requirement that the Bondi
light cones coincide, at $\cal I^+$, with the light cones emanating
from the world line of the center of mass of the particles' system.
This gauge was dubbed ``intrinsic'' because it is attached to the two-body dynamics. 
It was then suggested that
when a  quantity, such as the mechanical angular
momentum of the binary, is calculated by working with the dynamical equations of motion in the
center of mass frame, its value should correspond to the one obtained using  the intrinsic gauge.
The Bini-Damour formula  \cite{Bini:2012ji} being itself derived by this kind of calculations,
the relevant angular momentum loss to be used there
should then be that of the intrinsic --rather than the canonical--
gauge. And indeed  Damour's shear tensor (see (\ref{Dshear}) below) is precisely the one
obtained in the intrinsic gauge  \cite{Veneziano:2022zwh}. 

The flux of angular momentum in the intrinsic gauge gives a loss of mechanical $J$ which is larger, by an order in $G$, than the radiated $J$ since a large part of the initial $J$ is transferred to the non-radiative (static) part of the gravitational field. The above reasoning was verified \cite{Veneziano:2022zwh, Riva:2023xxm} to be correct when dealing with a two-body collision at the first non-trivial order in $G$. It was also given a BMS-invariant formulation in \cite{{Javadinezhad:2022ldc}}. Finally, it was possible to reproduce the static contribution to the loss of $J$ by amplitude techniques by suitably interpreting the role of``zero-frequency" gravitons \cite{Manohar:2022dea, DiVecchia:2022owy}.  

 Nonetheless, some doubts remained: it was far from obvious whether and how one could extend the definition of the intrinsic gauge to higher orders in $G$ or, even worse, whether any such gauge existed beyond the leading order. In the rest of this subsection I will present my understanding of recent attempts to clarify this point \cite{Ashtekar:2019rpv} \cite{Compere:2019gft}, \cite{Riva:2023xxm}, \cite{Chen:2021szm}, \cite{Chen:2021kug}, \cite{Mao:2023evc} \cite{Javadinezhad:2023mtp} \cite{Heissenberg:2024umh}.

Let us rewrite the claim of  \cite{Veneziano:2022zwh} about the total radiated angular momentum in the following suggestive form:
\bea
\label{totradJ}
&\Delta J^{(rad)}  = J_B(C_{AB}( - \infty)=0, u = + \infty) - J_B(C_{AB}( - \infty)=0, u = - \infty)  \\
& = \frac18 \int_{-\infty}^{+\infty} du \int d^2 \Omega Y^A\!\left( N^{BC} D_A C_{BC}-2 D_B(N^{BC} C_{AC}) + \frac12 D_A (N^{BC} C_{BC})  - \frac12 u D_A X \right )^{(CG)} \, , \nonumber \eea
where $(CG)$ means to stick to the Bondi gauge where $C_{AB}(u = - \infty)=0$. Note that, because of the intervening gravitational radiation, in general $C_{AB}(u = + \infty) \ne 0$ in the CG. However, if we want to compute the loss of the intrinsic angular momentum of the binary system we have to compare the initial ADM angular momentum with the final ADM angular momentum (i.e. with $J_{ADM}$ at $u \ra + \infty$). Following the arguments in \cite{Ashtekar:2019rpv},\cite{Ashtekar:2023wfn}, \cite{Ashtekar:2023zul} (see also  \cite{Riva:2023xxm}, \cite{Javadinezhad:2023mtp}) the final intrinsic angular momentum has to be computed in a Bondi frame which is the appropriate one at $u \ra + \infty$\footnote{In order to carry out this idea in all rigour it could be important to blow up the  point $u = + \infty$ and separate it from future time-like infinity $ i^+$. A step in this direction was recently made in \cite{Compere:2023qoa}. Another powerful approach, very much in the spirit of ADM, consists in adopting a Hamiltonian formalism  \cite{Henneaux:2019yax, Fuentealba:2022xsz, Fuentealba:2023syb} in which both spacelike and null infinity can be shown to possess an enlarged  BSM algebra containing ``logarithmic supertranslations". The connection between ADM angular momentum at $i^0$ and Bondi  angular momentum at $u = - \infty$ in the canonical gauge can then be established on solid algebraic grounds.}. In particular, it should satisfy two requirements: 

\begin{itemize}
\item It should be shear-free at $u = + \infty$ and therefore cannot be identified with the previously defined canonical gauge (which is shear-free at $u = - \infty$). As stressed in \cite{Ashtekar:2019rpv}  that means that the associated Poincar\'e subgroups of BMS at $u = \pm \infty$ are not the same. 
\item It should also involve a Lorentz boost ($\chi \ne 0$ in eq. (\ref{Lg})) in case the system is no longer in its original rest frame\footnote{ The gauge freedom associated with the translation subgroup (which determines the origin of the coordinate system) should also be carefully fixed in order to preserve Lorentz covariance of $J_{\mu\nu}$\cite{Riva:2023xxm, Javadinezhad:2023mtp, Heissenberg:2024umh}. Actually, the  discussion is somewhat subtler for the $J_{0i}$ Lorentz-boost charges, see  \cite{Heissenberg:2024umh} for a recent assessment.}.
\end{itemize}

Modulo this latter condition, that can be imposed separately (see e.g. \cite{Riva:2023xxm, Javadinezhad:2023mtp,  Heissenberg:2024umh}), one can then write \cite{{DiVecchia:2023frv}}:
\beq
\label{totJloss}
\Delta J^{(mech)} = J_B(C_{AB}(+ \infty)=0, u = + \infty)  -J_B( C_{AB}( - \infty)=0, u = - \infty) \,.
\eeq

It  is not at all obvious that a Bondi gauge  for which an expression similar to the last one in (\ref{totradJ}) exists for $\Delta J^{(mech)}$. In any case (\ref{totradJ})  and (\ref{totJloss}) imply that:
\bea
\label{Deltas}
& \Delta J^{(mech)} = \Delta J^{(rad)} + J_B(C_{AB}( + \infty)=0, u = + \infty)  - J_B(C_{AB}( - \infty)=0, u = + \infty) \nonumber \\
& \equiv \Delta J^{(rad)} + \Delta J^{(stat)}\, .
\eea
The above definition of $\Delta J^{(stat)}$ can be written in the form:
\bea
\label{DJstat}
& \Delta J^{(stat)} = J_B(C_{AB}( - \infty)= - S_{AB}, u = + \infty)  - J_B(C_{AB}( - \infty)=0, u = + \infty) \, , \nonumber \\
& S_{AB} \equiv \int_{- \infty}^{+ \infty} d u N_{AB} =  C_{AB}( + \infty) - C_{AB}( - \infty)\,, 
\eea
where $S_{AB}$ is the so-called memory (including its non-linear part), a BMS-invariant observable. The name ``static"  stems from the fact that a non vanishing memory (needed for a non-vanishing $\Delta J^{(stat)}$) is tied to the presence of a $\omega^{-1}$ singularity in the Fourier transform of $C_{AB}(u)$ \cite{Strominger:2014pwa} while it is  insensitive to the finite-frequency modes corresponding to physical on shell gravitons. In terms of scattering amplitudes it is captured by suitably treating ``zero-frequency gravitons" \cite{DiVecchia:2022owy, DiVecchia:2023frv}. It is clear, both from the memory point of view and from Feynman diagram considerations, that $\Delta J^{(stat)}$ does have contributions starting at ${\cal O}(G^2)$. By contrast $\Delta J^{(rad)}$, as defined in  (\ref{totradJ}),  can only start at ${\cal O}(G^3)$.

The bottom line is that, while the two components of the angular momentum loss are both to be computed starting  from the canonical Bondi gauge at $u = - \infty$, only the radiative part of the loss sticks to that gauge at all $u$, while the static and total loss involve different initial and final Bondi frames adapted to the initial and final two-body system, respectively. It is quite clear however that both $\Delta J^{(rad)}$ and  $\Delta J^{(stat)}$, in spite of not being written in BMS-invariant form, have a BMS invariant meaning by referring to some uniquely specified Bondi frames.
This is my simple take from the arguments given in  \cite{Ashtekar:2019rpv}, \cite{Compere:2019gft}, \cite{Riva:2023xxm}, \cite{Chen:2021szm}, \cite{Chen:2021kug}, \cite{Mao:2023evc}, \cite{Javadinezhad:2023mtp}, \cite{Compere:2023qoa}  to which I refer the reader  for all the (very essential) details\footnote{In several of these papers gauge invariant expressions are given for the Lorentz charges by adding terms controlled by $C_{AB}(u)$ (see e.g. Appendix E of \cite{Compere:2023qoa}). They reduce to our gauge-fixing proposal at $u = \pm \infty$ since the additional terms vanish by virtue of our gauge choices.}.

To conclude, we have seen how shear and news tensors play a fundamental role in describing both dynamical and static properties of the gravitational field at large distances. News, as well as shear variations in time,  are  BMS invariant. They are related to physical effects such as gravitational radiation and memory, the latter being in turn related to soft theorems \cite{He:2014laa,Strominger:2014pwa}.
On the other hand,  shear itself is not BMS invariant. Its fixing appears to play an important  role in connecting ADM to BMS quantities --at least in asymptotic non-radiative regions--  and in  distinguishing quantities carried by gravitational waves from those belonging to the material system emitting them. 
 Although several technical issues (like the distinction between BMS charges and the generators of BMS transformations) remain to be fully clarified, I believe that the important issue of how the total initial angular momentum splits itself unambiguously among the binary system, gravitational static field, and radiation in the final state is finally on the way of being fully clarified through  a BMS-ADM connection.

\section*{Acknowledgements}

I am grateful to Abhay Ashtekar, Thibault Damour, Marc Henneaux, Eric Poisson, Massimo Porrati, Chia-Hsien Shen, and Grisha Vilkovisky for many discussions on the topic of angular momentum loss, and to Thibault Damour, Robert Wald, and Alexander Zhiboedov, for useful comments and correspondence on the content of the Appendix. I am also most grateful to Paolo Di Vecchia, Carlo Heissenberg, and Rodolfo Russo for a long and most enjoyable collaboration on all these topics and for comments on a preliminary version of this paper.

\appendix

\section{The puzzle of local energy conservation}

   Consider  the simple case of the gravitational collision of two massless (scalar for simplicity) particles al large impact parameter.  Let us  rewrite the first of (\ref{eveq}) and its analog at past null infinity:
\bea
 &G \partial_u  {\cal  M}_{+}    =  - \frac{1}{8} N_{+AB} N_+^{AB}  +  \frac14 D_AD_B  N_+^{AB} - 4 \pi G r^2 T_{uu} (u, \phi^A) \, , \nonumber \\
 & G \partial_v   {\cal M}_{-}    =  +  \frac{1}{8} N_{-AB} N_-^{AB} -  \frac14 D_AD_B  N_-^{AB} + 4 \pi  G r^2 T_{vv} (v, \phi^A) \, , 
 \label{pastdmasp}
	\eea
where we have reinserted Newton's constant and added  possible matter  contributions  with asymptotic stress-energy tensor $T_{\mu \nu}$.

Integrating both equations from $- \infty$ to $+ \infty$  gives the differences between each Bondi mass aspect at $\pm \infty$. However, under the assumptions (satisfied in our example) that: i) there is no energy in the system at $v= - \infty$ (i.e. no initial massive particle), and: ii) there is no energy remaining in the system at $u = + \infty$ (no massive final particle, in particular  no black hole is formed)  we can set ${\cal M}_{-}(- \infty, \phi^A) = {\cal M}_{+}(+\infty, \phi^A) =0$ and rewrite (\ref{pastdmasp}) as
\bea
 &{\cal M}_{+}(-\infty, \phi^A) - {\cal M}_{-}(+ \infty, \tilde{\phi}^A)  =  
 \int_{- \infty}^{+ \infty}  du \left( 4 \pi r^2 T_{uu} + \frac{1}{8G} N_+^{AB} N_{+AB} - \frac{1}{4G} D_AD_B N_+^{AB}\right)(u, \tilde{\phi}^A) \nonumber \\
 &-  \int_{- \infty}^{+ \infty}  dv \left(4 \pi r^2  T_{vv} + \frac{1}{8G} N_-^{AB} N_{-AB} - \frac{1}{4G} D_AD_B  N_-^{AB} \right) (v, \phi^A)\, .
   \label {Relation}
    \eea
 The`` local conservation equation" of \cite{Strominger:2013jfa} is the statement that the l.h.s. of (\ref{Relation}) vanishes if  $\tilde{\phi}^A = \bar{\phi}^A$ with  $\bar{\phi}^A$ and $\phi^A$ antipodally related ($\bar{\theta} = \pi - \theta, \bar{\phi} = \pi + \phi$), implying:
\bea
&\int_{- \infty}^{+ \infty}  du \left( 4 \pi r^2  T_{uu} + \frac{1}{8G} N_+^{AB} N_{+AB} - \frac{1}{4G} D_AD_B N_+^{AB}\right)(u, \bar{\phi}^A) \nonumber \\
 &=  \int_{- \infty}^{+ \infty}  dv \left(4 \pi r^2  T_{vv} + \frac{1}{8G} N_-^{AB} N_{-AB} - \frac{1}{4G} D_AD_B  N_-^{AB} \right) (v, \phi^A)\, .
  \label{Fluxrelation}
 \eea

At leading order in $G$ the two particles deflect each other by a small angle $\theta_s$ in the direction of the impact parameter. As a result, $T_{uu}(\bar{\phi}^A)$ and $T_{vv}(\phi^A)$  do {\it not}  coincide at ${\cal O}(G)$. Furthermore, the terms bilinear in the news are of higher order in $G$.   The only possibility to recover (\ref{Fluxrelation}) at ${\cal O}(G)$ is that the last  terms on both sides of (\ref{Fluxrelation}) (which unlike the gravitational radiation terms already occur at ${\cal O}(G)$) compensate for  the above mismatch. 

 Let us check that this does indeed happen. Equation 
 (\ref{Fluxrelation}) simplifies further in our case, since the initial news functions $N_-^{AB}$ vanish, and at ${\cal O}(G)$ becomes:
\bea
&\int_{- \infty}^{+ \infty}  du \left( 4 \pi r^2  T_{uu}  \right)(u, \bar{\phi}^A)  - \int_{- \infty}^{+ \infty}  dv \left(4 \pi r^2  T_{vv} \right) (v, \phi^A)  = \int_{- \infty}^{+ \infty} du \frac{1}{4G} D_AD_B N_+^{AB} (u, \bar{\phi}^A) \nonumber \\
& =  \frac{1}{4G} \left( D_AD_B  C_+^{AB}(u= {+ \infty}, \bar{\phi}^A)-  D_AD_B   C_+^{AB}(u= {-\infty}, \bar{\phi}^A)\right)
\, .
   \label{OG2}
 \eea

   The lowest order shear $ C_+^{AB}(u= {-\infty})$ corresponding to two incoming particles with  mass $m$ and arbitrary center-of-mass  velocity $v$ was already computed in \cite{Damour:2020tta}\footnote{ While the result in \cite{Damour:2020tta} depends  on the (Bondi) gauge  used the difference  in (\ref{OG2}) is gauge- independent.}. In the center of mass system it reads:
  \beq
  \label{Dshear}
   C_+^{AB}(u= {-\infty}) = \frac{ 4 G E}{ (v^{-2} - \cos^2 \theta)} {\rm diag} (\sin^2 \theta, -1) \, ,
  \eeq
  where we are using polar coordinates with the initial particles coming from the two poles of the celestial sphere.
  
The double covariant divergence of  $C^{AB}$ yields:
  \beq
 \sqrt{\gamma} D_A D_B C_+^{AB}(u= {-\infty},\theta, \phi) = 4 G E \left[ \partial_{\theta}^2 \left( \frac{\sin^3 \theta}{(v^{-2} - \cos^2 \theta)}\right) + \partial_{\theta} \left( \frac{\sin^2 \theta \cos \theta}{(v^{-2} - \cos^2 \theta)}\right) \right] \, .
  \label{ddiv}
  \eeq
 As expected this quantity vanishes upon full integration over the sphere.  The only subtle point is to take carefully the massless limit while preserving this property.  By doing so one finds \cite{GVunp}\footnote{It is easy to check the result (\ref{delta-}) numerically by plotting the function  (\ref{ddiv}) for smaller and smaller  $m$.}:
\beq
\frac{1}{4G} D_AD_B  C_+^{AB}(u= {-\infty}, {\bar \phi}^A) \rightarrow 4  \pi  E~ \left(  \delta^{(2)}(\bar{\Omega} - \Omega({\overrightarrow p})) +  \delta^{(2)}(\bar{\Omega} - \Omega({-\overrightarrow p})) - \frac{1}{2 \pi} \right)\, ,
\label{delta-}
\eeq
where $\Omega({\pm \overrightarrow p})$ represents the direction of the momenta $\pm \overrightarrow p$ of the two incoming particles and, in standard spherical coordinates,
\beq
           \delta^{(2)} (\Omega - \Omega') = \delta (\cos \theta- \cos \theta')  \delta (\phi - \phi') \, ; \,   \int d^2 \Omega ~ \delta^{(2)} (\Omega - \Omega') = \frac{1}{4\pi} \, .
\label{deltagen}
\eeq
 As a consistency check,  (\ref{delta-}) integrates to zero  as it should.  
 Similarly, at $u= + \infty$ we find:
\beq
\frac{1}{4G} D_AD_B  C_+^{AB}(u= {+\infty},{\bar \phi}^A) = 4  \pi  E~  \left(  \delta^{(2)}(\bar{\Omega} - \Omega({\overrightarrow p '})) +  \delta^{(2)}(\bar{\Omega} - \Omega({-\overrightarrow p '})) - \frac{1}{2 \pi} \right)\, .
\label{delta+}
\eeq

Note that each quantity  in (\ref{delta-}) and (\ref{delta+}) is of ${\cal O}(1)$. 
However, their difference, as it appears in (\ref{OG2}), is of ${\cal O}(G)$ since the constants cancel and $\overrightarrow p '$ differs from $\overrightarrow p$,  by the ${\cal O}(G)$ scattering angle $\theta_s$.
Finally, the two  $\delta$-function terms in (\ref{delta-}) precisely cancel the contribution from  $T_{vv}$ in  (\ref{OG2}) and
  the  two $\delta$-function terms in (\ref{delta+}) cancel the contribution from $T_{uu}$ in (\ref{OG2}). These latter contributions are impulsive in nature since they correspond to two particles coming in at a specific value of $v$ and coming out at a (time delayed) specific value of $u$. Although a cancellation does occur after integration over $v$ and $u$, I see no reason to associate any observable energy flow with the r.h.s. of (\ref{OG2}).

The above result can be easily generalized to the case of many initial particles scattering at ${\cal O}(G)$ \cite{Veneziano:2022zwh}. Each particle (labelled by $a$) in the initial state will contribute a term:
\beq
\frac{1}{4G} D_AD_B    C_{+,a}^{AB}(u= - \infty, {\bar \phi}^A) = 4  \pi  E_a~\left(  \delta^{(2)}(\bar{\Omega} - \Omega({\overrightarrow p_a})) - \frac{1}{4 \pi} \right) = 4 \pi \int dv r^2 T^{a}_{vv} - E_a \, ,
\label{delta+-}
\eeq
while each particle in the final state will contribute a term:
\beq
\frac{1}{4G}D_AD_B   C_{+,a}^{AB}(u= +\infty,{\bar \phi}^A) = 4  \pi  E_a~ \left(  \delta^{(2)}(\bar{\Omega} - \Omega({\overrightarrow p_a'})) - \frac{1}{4 \pi} \right) = 4 \pi \int du  r^2 T^{a}_{uu} -E_a \, .
\label{delta++}
\eeq
This makes the generalization to the case of many massless colliding particles straightforward since their contribution to the lowest order shear is additive  \cite{Veneziano:2022zwh}. 
We therefore conclude that equation (\ref{Fluxrelation}) is indeed valid so that ${\cal M}_{+}(-\infty, \bar{\phi}^A) = {\cal M}_{-}(+ \infty, \phi^A) $.

 As I have already mentioned, the above matching at $i^0$ has been proven formally in much greater generality (see e.g.\cite{Satishchandran:2019pyc, Prabhu:2019fsp}).
Nonetheless, the explicit calculation reported here strongly suggests  that a calorimeter will fail to find that the time-integrated energy coming in at each angle comes out at the corresponding antipodal angle\footnote{This conclusion is in agreement with the more general arguments given recently in \cite{Rignon-Bret:2024mef}.}.  At the order at which we have been working only the scalar particles bring in and carry out energy and there is no antipodal connection between the  angles at which they do so. The fact that shear makes the two mass aspects coincide locally looks to be unrelated to an actual flow of gravitational-radiation energy coming out at each angle as recently computed perturbatively in $G$ in \cite{Herrmann:2024yai}: the change in the shear is rather there to insure that the BS metric functions, as one approaches $i^0$ from ${\cal I}^-$ and ${\cal I}^+$,  both agree with the corresponding ones in the ADM formalism.

\end{document}